# THE FLOWING SYSTEM GASDYNAMICS
## Part 2: Euler's momentum conservation equation solution

S.L. Arsenjev, I.B. Lozovitski[1], Y.P. Sirik

*Physical-Technical Group*
*Dobroljubova street 2, 29, Pavlograd, Dnepropetrovsk region, 51400, Ukraine*

The solution of a momentum conservation equation for the gas and liquid stream in the flowing element is obtained on the basis of the modern approach to a problem on contact interaction of bodies and mediums. A flowing element, system are: pipe, tube, orifice, mouthpiece, diffuser, etc. and its combination. The integration of the differential equation has reduced to distribution law of static head along the length of flowing element and has proved the elementary algebraic solution that introduced in the previous paper by these authors. The received solution allows to describe the motion of fluid medium in non-stationary conditions, under action of any time-varying physical factors: a roughness of streamlined surface, the area of the section of the flowing element, the heat exchange with the streamlined surface, the technical work, the additional weight flow of fluid medium.

## Nomenclature

| | |
|---|---|
| $\rho$ | mass density of fluid medium |
| $\lambda$ | coefficient of hydraulic friction |
| $l$ | current length of stream |
| $L$ | general length of flowing element |
| $\bar{L}$ | general caliber length of flowing element, $L/D$ |
| $D$ | internal diameter of flowing element |
| $\zeta_{in}$ | coefficient of local hydraulic resistance for inlet into flowing element |
| $\zeta_{ex}$ | coefficient of local hydraulic resistance for outlet from flowing element |
| $p_0, p_h$ | quantities of pressure on inlet and outlet of flowing element accordingly |
| $p$ | static head in stream |
| $V$ | stream velocity determined by weight flow |

## 1 Introduction

The problem of the solution of the momentum conservation equation for stream of fluid medium in the flowing element has occurred from the moment of its publication by Euler in 1755. This problem is bound with question: what its solution means? And this problem, in turn, has demanded clearing up of the physical meaning contained in the given equation. If the equation describes motion of an ideal fluid medium as Euler and his following considered, then there is question: what is physical reason of the change of the relation between two components of equation in its left-hand side with conservation of equality of its total to zero value? Apparently, the medium which is not having the inside structural kinematic and dynamic linkage can not change parameters of motion. The stream of the fluid medium which is not possessing the above mentioned property, can be considered only as a whole with stationary values of motion parameters in all its parts by analogy with the Newton's first law. All the more, such medium is not capable to interact by contact with an environment. It is impossible neither to understand sense of Euler's equation, nor to decide its productively owing to the physical insignificancy of such medium, termed ideal fluid.

In the XX century, the attempt of substitution in the left-hand side of Euler's equation of the static head losses along the length of pipe in accordance of Weissbach-Darcy's formula was undertaken after experimental determination of allocation of static head in fluid flow in pipe. This means should ensure the conversion to consideration of an real fluid medium possessing such inside structural property as viscosity and capable to response to external actions on thought of the hydrodynamic scientists of the XX centuries. However and this means was unproductive. The introduction in Euler's equation of the third member, nonzero, physically means what alongside with kinetic and potential energy in the stream of fluid medium is contained a certain

---

[1]Corresponding author.
Phone.: (38 05632) 38892, 40596
E-mail: loz@inbox.ru

third type of energy that is related to the contact interaction of the stream with a streamlined surface. Euler's equation has ceased to be the equation as such both physically and mathematically in this form. Thus, the problem of productive usage of Euler's equation as one of the basic equations for exposition of fluid medium dynamics is, that during its solution:

− to retain its form of record;
− to take into account mathematically correctly influence of contact interaction on a relation of kinetic and potential energy, which varies along the length of stream but, the aggregate content of this two types of energy is remained invariable;
− to know that the solution should give the potential distribution, that is the static head in stream along the length of flowing element according to Torricelli-Galilei-Borda-Du Buat (TGBD) relationship for free fall and Bernoulli's principle about correspondence of relation of kinetic and potential energy of fluid flow in the pipe to this relation for free fall of solid in the gravitational field.

Such solution of the momentum conservation equation should have an as much as possible generalized form for stream of gas medium with the particular form for liquid. At the same time both solutions will appear fundamental for gas dynamics and hydrodynamics accordingly.

## 2 Euler equation solution

So, after almost 250 years the Euler's partial differential equation looks like:

$$V(l)\frac{\partial V(l)}{\partial l} + \frac{1}{\rho(l)}\frac{\partial p(l)}{\partial l} = 0 \qquad (1)$$

Basing on the conception about contact interaction of fluid flow with the streamline surface, it is necessary to write up the equation (1) in the form:

$$d[\rho(l)V^2(l) + p(l)] = 0, \qquad (2)$$

where $p$ is static head distributed along the length of the pipe proportionally to intensity of contact interaction of stream with wall.

Out of the equation (2) follows:

$$\rho(l)V^2(l) + p(l) = const \qquad (3)$$

The value of the constant is determined out of requirements $V=0$, $p=p_0$, with allowance for which one the equation (3) acquires particular form:

$$\rho(l)V^2(l) + p(l) = p_0 \qquad (4)$$

The intensity of contact interaction of the stream with the wall is expressed by dependence:

$$dp(l) = \lambda \frac{1}{D}\frac{\rho(l)V^2(l)}{2}dl \qquad (5)$$

what will match the differential form of writing of Weissbach-Darcy's formula for the stream of the fluid medium in pipe. The integration of the expression (5) gives the allocation of static head along the length of pipe accordingly to the intensity of contact interaction of stream with wall in the form:

$$p(l) = \lambda \frac{1}{D}\int_0^l \frac{\rho(l)V^2(l)}{2}dl + C \qquad (6)$$

where an actual longitudinal coordinate, $l$ needs to be counted out from the outlet section of the pipe. The integration constant $C$ in the equation (6) is determined out of conditions $l = 0$, $p = p_h$. In result, $C = p_h$ and equation (6) acquires the form:

$$p(l) = \frac{\lambda}{D}\int_0^l \frac{\rho(l)V^2(l)}{2}dl + p_h \qquad (7)$$

In result of substitution (7) in (4), and taking into account $\rho(l)V(l) = const$, we now have:

$$\frac{\rho(l)V^2(l)}{2} = \frac{p_0 - p_h}{2 + \frac{\lambda}{D}\rho(l)\int_0^l \frac{1}{\rho(l)}dl} \qquad (8)$$

in turn, having substituted (8) in (7), we receive the distribution law of the static head along the length of the flowing element:

$$p(l) = (p_0 - p_h) \times$$
$$\times \frac{\frac{\lambda}{D}\rho(l)\int_0^l \frac{1}{\rho(l)}dl}{2 + \frac{\lambda}{D}\rho(l)\int_0^l \frac{1}{\rho(l)}dl} + p_h \equiv$$
$$\equiv (p_o - p_h) \times$$
$$\times \frac{\frac{\lambda}{D}\rho(l)\int_0^l \frac{1}{\rho(l)}dl}{1 + \frac{\lambda}{D}\rho(l)\int_0^l \frac{1}{\rho(l)}dl + \varsigma_{in} + \varsigma_{ex}} + p_h \qquad (9)$$

The obtained expression determines allocation of the static head along the length of the flowing element as applied to gas medium.

For liquid stream in pipe, it is valid $\rho(l) \cdot V^2(l)/2 = const$ and the expression (8) will look like:

$$\frac{\rho V^2}{2} = (p_0 - p_h)\frac{1}{2 + \frac{\lambda}{D}L} \qquad (10)$$

Substituting (10) in (7), we obtain a particular solution for liquid:

$$p(l) = (p_0 - p_h)\frac{\lambda\frac{l}{D}}{2 + \lambda\overline{L}} + p_h \equiv$$
$$\equiv (p_0 - p_h)\frac{\lambda\frac{l}{D}}{1 + \lambda\overline{L} + \varsigma_{in} + \varsigma_{ex}} + p_h \qquad (11)$$

## 3 Concluding remarks

Thus, later almost 250 years, the physically adequate and the mathematically correct solution of the equation is obtained which is one of the fundamental relationship in the field of gas dynamics and hydrodynamics. The solution of momentum conservation equation for stream of fluid medium in flowing element in differential form match to its elementary algebraic solution [1] what confirms the uniqueness of the solution. At the same time the differential approach has allowed to receive the solution suitable for description of motion of fluid medium in non-stationary conditions, under action of any time-varying physical factors: a roughness of streamlined surface, the area of the section of the flowing element, the heat exchange with the streamlined surface, the technical work, the additional weight flow of fluid medium.

---